\documentclass[review]{elsarticle}

\journal{Journal of \LaTeX\ Templates}

\usepackage{amsmath}
\usepackage{amssymb,amsfonts,latexsym}
\usepackage{bm}
\usepackage[mathcal]{euscript}
\usepackage{graphicx}
\usepackage{epsfig}
\usepackage{color}
\usepackage{pifont}
\usepackage{float}
\usepackage{gensymb}
\usepackage{color}

\definecolor{blue}{rgb}{0,0,1}
\definecolor{bleuf}{rgb}{0,0,0.9}
\definecolor{rougef}{rgb}{0.9,0,0}
\definecolor{green}{rgb}{0,0.5,0}
\definecolor{red}{rgb}{1,0,0}
\definecolor{pink}{rgb}{0.9,0.3,0.7}
\definecolor{azur}{rgb}{0,0.5,0.5}
\definecolor{orange}{rgb}{1,0.5,0.2}
\definecolor{brown}{rgb}{0.5,0,0}

\newcommand{\be}{\begin{equation}}
\newcommand{\ee}{\end{equation}}
\newcommand{\ben}{\begin{equation*}}
\newcommand{\een}{\end{equation*}}
\newcommand{\ba}{\begin{eqnarray}}
\newcommand{\ea}{\end{eqnarray}}

\begin{document}

\begin{frontmatter}

\title{Mode I fracture of a biopolymer gel: rate-dependent dissipation and large deformations disentangled}

\author{Maxime Lefranc $^{\star}$}
\address{PSL Research University, ESPCI-Paris Tech, PSL$^{\star }$, UMR Gulliver, EC2M, 10 rue Vauquelin, 75231 Paris Cedex 05, France}

\cortext[mycorrespondingauthor]{Corresponding author}
\ead{maxime.lefranc@espci.fr}

\author{Elisabeth Bouchaud}
\address[mymainaddress]{PSL Research University, ESPCI-Paris Tech, PSL$^{\star }$, UMR Gulliver, EC2M, 10 rue Vauquelin, 75231 Paris Cedex 05, France}
\address[mysecondaryaddress]{CEA-Saclay, IRAMIS, SPEC,  F-91191 Gif-sur-Yvette Cedex, France}

\begin{abstract}

We have designed a new experimental setup able to investigate fracture of soft materials at small scales. At high crack velocity, where energy is mostly dissipated through viscoelastic processes, we observe an increasingly large high strain domain in the crack tip vicinity.
Taking advantage of our ability to determine where linear elasticity breaks down, we derive a simple prediction for the evolution of the energy release rate with the crack velocity. 

\end{abstract}

\begin{keyword}
\texttt{Fracture, Biopolymer, Gel, Large deformation, Rate-dependent dissipation}
\end{keyword}

\end{frontmatter}

\section{Introduction}

Soft materials are of great interest both for industrial and fundamental reasons. Materials with shear moduli in the range 0.1-100 kPa can be found indeed in very different applications, from tissue engineering to food industry or geophysics. While ubiquitous, their mechanical properties are poorly understood, especially when related to fracture. First, it is an experimental challenge to perform controlled fracture experiments on such materials, which are difficult to grip~\cite{Baumberger2006a} and which, sometimes, flow at a macroscopic scale~\cite{Foyart2013}. Second, the fracture mechanics of these materials is made particularly complex by  the interplay of  large deformations~\cite{Livne2013} and viscoelastic processes ~\cite{Baumberger2009}.

Fracture of soft materials has been widely studied both theoretically~\cite{Hui1992, Persson2005, Bouchbinder2011a} and experimentally~\cite{Suo2012,Baumberger2006a, Properties2000, Tanaka2000, Seitz2009}. Large efforts were made  to measure the rate dependency of the energy release rate~\cite{Baumberger2006a, Tanaka2000, Seitz2009} and to model it using either viscoelastic models~\cite{Persson2005, Hui1992} or microscopic damage models \cite{Baumberger2009}. However, in very deformable materials, microscopic cohesive zone models deal with what happens in the very non linear vicinity of the crack tip \cite{Livne2013, Hui2003}. Because non linear processes will change the way the elastic energy is conveyed toward the crack tip and then dissipated in its vicinity, it is crucial to characterize crack propagation at both macroscopic and microscopic scales. Biopolymer physical gels are good candidates to gain a deeper understanding of the fracture properties of soft materials. Indeed, linear and nonlinear mechanics of these gels have been widely characterized \cite{Labropoulos2001, Hall2010}. 


In this Letter, we introduce a new experimental setup that was designed  to probe the mechanical response of soft materials in the vicinity of the crack tip.  As described in Section \ref{sec2}, Mode I cracks are grown at controlled velocity $V$ in a 2D sample of agar gel. In Section \ref{sec3}, we analyse crack profiles when varying $V$. We show that Linear Elastic Fracture Mechanics (LEFM) breaks down at a distance from the crack tip which increases with $V$ and which is proportional to the viscoelastic dissipated energy. Finally, Section \ref{sec4} is devoted to a discussion of the experimental results.

\section{Material and Methods}\label{sec2}

\paragraph{Material and sample preparation}

 Agar gels are prepared by dissolving agar powder (Sigma Aldrich, average molecular weight $120kg.mol^{-1}$, chain contour length $\Lambda=200 nm$) in hot water. A 1.5\%wt solution of agar is stirred for 1h at 95 C without any water loss. The obtained sol is then kept at 70 C and used within a day. The gelation mechanism is simple: at temperatures lower than 37 C,  dissolved chains start interacting via H bonds and locally form junctions, leading to the formation of a weak physical network.

Linear dynamical rheology measurements are performed at 22 C (Anton Paar Rheometer, 1\% strain in the linear viscoelastic range, roughened cone and plate geometry), showing that both $G'$ and $G"$ are nearly constant within the frequency range (0.1 Hz-100 Hz), with $G"$ being 40 times smaller than $G'$ ($G'$=20kPa). The material being incompressible, its Young modulus is $E=3G'$=60kPa. Knowing $E$ gives us an order of magnitude for the typical distance between the junctions $\xi=(kT/E)^{1/3} \simeq 10 nm$. The non linear elasticity of agar gels was investigated by Pavan \& al \cite{Hall2010}. This work shows that agar gels are not neo-Hookean but would rather exhibit exponentially stiffening. 

Just before casting the gel, 1ml of sol is dyed with 2.5$\mu$l of a dispersion of carbon black in indian ink (inset of Fig. \ref{crackzoomlent}).

\paragraph{Fracture experiments}

\begin{figure}[H] 
\center
\includegraphics[width=0.5\columnwidth,height=0.33\columnwidth]{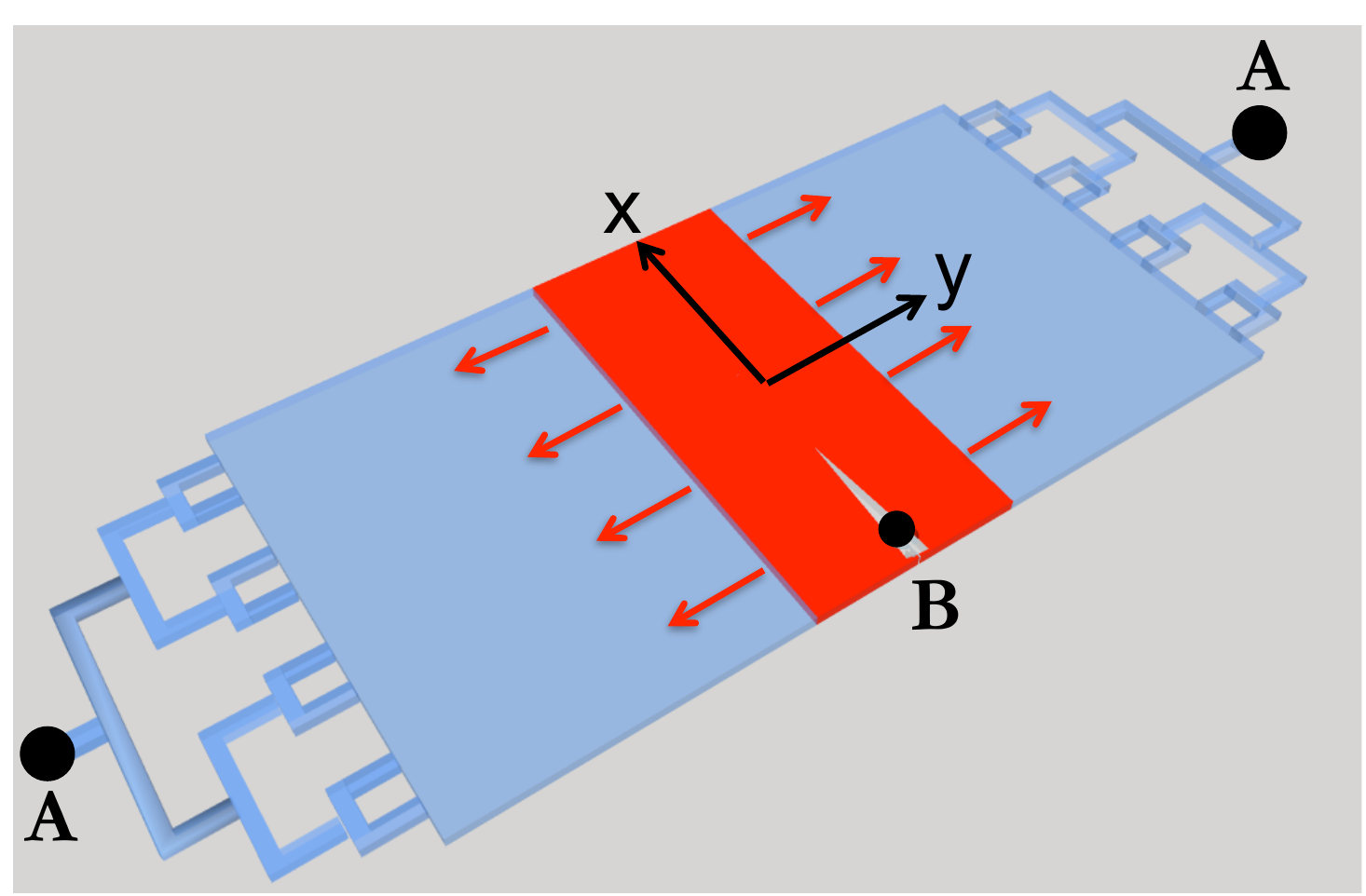}
\caption{{\bf On a chip mechanical test cell.} Hele-Shaw chamber containing the sample of gel (in red) surrounded by fluorinated oil (in blue). Oil is sucked out of the chamber from points A, which results in a displacement of the gel boundaries in the y-direction. In this plane stress configuration, a mode I crack grows in the x-direction.}

\label{setup}
\end{figure}

Our new device is composed of a rigid chamber made of two glass plates of thickness 1mm, separated by a distance of 350$\mu $m (Fig.~\ref{setup}).  It is microfabricated in a clean room, using microfluidic stickers technology \cite{Bartolo2008}. The edges of the chamber are sealed with photocurable glue.

It is first filled with FC3280 fluorinated oil (in blue in Fig.~\ref{setup}). The hot sol (in red in Fig.~\ref{setup}) is injected in the middle of the chip through a hole B in the upper plate, and cooled down to 22C for 1 hour until complete gelation. Before starting the fracture experiment, the gel sample dimensions are 0.35mm (thickness) in the z-direction, 25mm in the x-direction and 14mm in the y one (Fig. \ref{setup}). To assure perfect slippage and prevent adhesion of the gel on all the walls of the chamber, the latter were coated with an acrylamide brush using a two-step surface reaction \cite{Lefrancthese}. Two synchronized syringe pumps connected to points A (Fig.~\ref{setup}) suck the oil out of the chamber at a prescribed flow rate $Q$. 

The oil being immiscible with the gel, the interfaces between the fluid and the material remain sharp. Acting as a tensile machine of low stiffness, the oil outflow imposes a velocity of displacement proportional to $Q$ to these two boundaries. However, due to the low rigidity of the tensile system, the displacement along the oil/gel interface is non uniform as soon as a crack starts propagating.

Under deformation, a single mode I crack is nucleated at point B from a built-in notch of length 5mm and radius of curvature at the tip $\simeq 10\mu m$. The crack tip is observed under a microscope with a 4$\times $ magnification (observation field 9 $mm^2$, pixel size 1.5$\mu$m). 

B is connected to a water reservoir so that the propagating crack is filled with water instead of air. As a consequence, our gel being mostly made of water, we suppress surface tension forces which might generate a spurious blunting of the crack tip, smoothing it at scales smaller than  the elastocapillary length $\gamma/E= 1.2\mu m$. 

\paragraph{Experimental observations}

At fixed $Q$, after a transient acceleration, the crack reaches a propagation regime where the crack velocity $V$ varies very slowly with time, until it starts feeling the edges of the specimen. Within the steady-state regime, sequences of images are captured and processed with a home-made image analysis routine detecting the crack lips. At this concentration, changing the control parameter $Q$ from 0.1 to 1000$\mu$L/min enables one to tune $V$ between 1$\mu$m.s$^{-1}$ and 1cm.s$^{-1}$: $V$ being always much smaller than the Rayleigh wave speed $V_R\simeq 1m.s^{-1}$, crack propagation can be considered as quasi-static. Note that for velocities lower than $1 \mu m.s^{-1}$, crack propagation becomes intermittent: under stress, junctions have time to unzip and the network to relax. For $V$ higher than 1 $cm.s^{-1}$, it becomes difficult to reach a steady state. 

For each experiment, the crack opening displacement (COD) is extracted from the observed shape of the crack, while the full displacement field around the crack tip is determined by using the CorreliQ4 Digital Image Correlation (DIC) code \cite{Hild2006}. Full DIC results will be reported elsewhere~\cite{Lefrancthese}, where the dominance of mode I over mode II ($K_I$=50 $K_{II}$) is shown. In this plane stress configuration, we can thus relate directly the energy release rate $\cal {G}$$= K_I^2/E$ \cite{Lawn} with the stress intensity factor (SIF) $K_I$. ${K_I}$ can indeed be extracted from the measurement of the displacement along the crack direction, behind the crack tip.

\section{Experimental results}\label{sec3}

\begin{figure}[H] 
\center
\vspace{0.0cm}
\includegraphics[width=0.75\columnwidth,height=0.56\columnwidth]{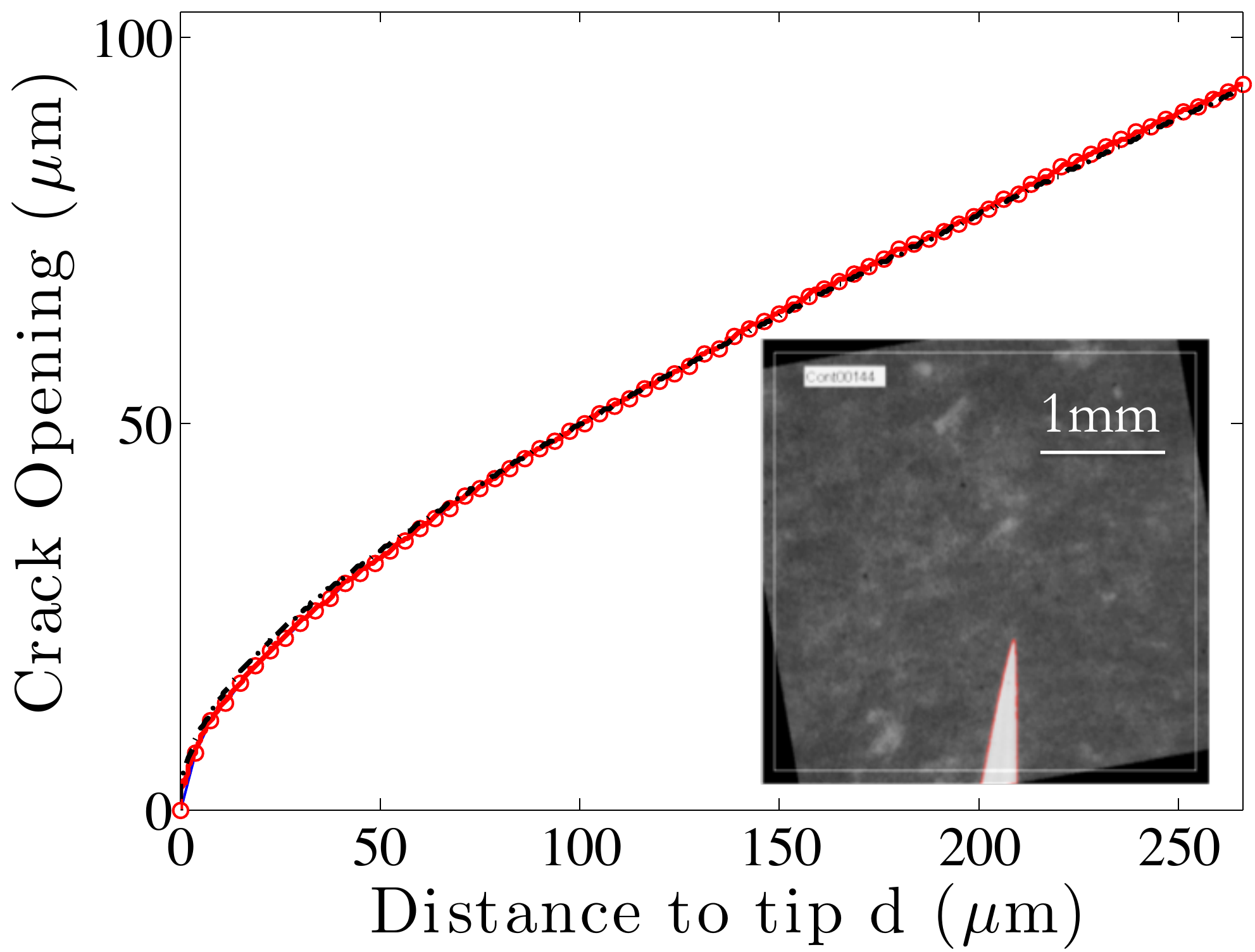}
\vspace{-0.5cm}
\caption{{\bf Slow crack ($V$=3$\mu$m.s$^{-1}$)}. Crack opening (displacement along the crack lips) as a function of the distance d to the crack tip. Red line with circles: experimental results. Black dashed line: fit with the LEFM expression \ref{u}.
{\it Inset}: Picture of a low-velocity crack (white) in a dyed 1.5$\%$wt agar gel. The interior of the crack is filled with water. The contour of the crack is superimposed in red. }
 
\label{crackzoomlent}
\vspace{-0.0cm}
\end{figure}

Fig.~\ref{crackzoomlent} shows a typical snapshot of a crack in the low velocity range ($V$=3$\mu$m.s$^{-1}$) and its corresponding COD. A video sequence can be accessed from the link provided in~\cite{SM1}. This profile fits in well the LEFM theory and Williams's prediction (1957) for the crack opening displacement $u$:

\begin{eqnarray}
u(d)={K_I \over E}\sqrt {8\over \pi} d^{1/2}\left[ 1+ {d\over d_{1}}+ \left( {d\over d_{2}}\right) ^{2} \right]
\label{u}
\end{eqnarray}
where $d$ is the distance to the crack tip, and $E$ is the gel's Young modulus. Fig.~\ref{crackzoomlent} shows a fit of the experimental profile based on the first three terms of Williams expansion. For this low-velocity crack, $d_{1}$ and $d_{2}$ are respectively  $900\mu m$ and $4mm$, showing that, even if they are essential to capture the crack profile, these two extra terms are only relevant far from the crack tip. This shows the existence of a zone of $K_I$-dominance at the crack tip, i.e. a region of space in which the Irwin $d^{1/2}$ term dominates \cite{Lawn}. Hence $K_I$  can be deduced directly from the COD measurement.

\begin{figure}[H] 
\center
\vspace{0.0cm}
\includegraphics[width=0.75\columnwidth,height=0.56\columnwidth]{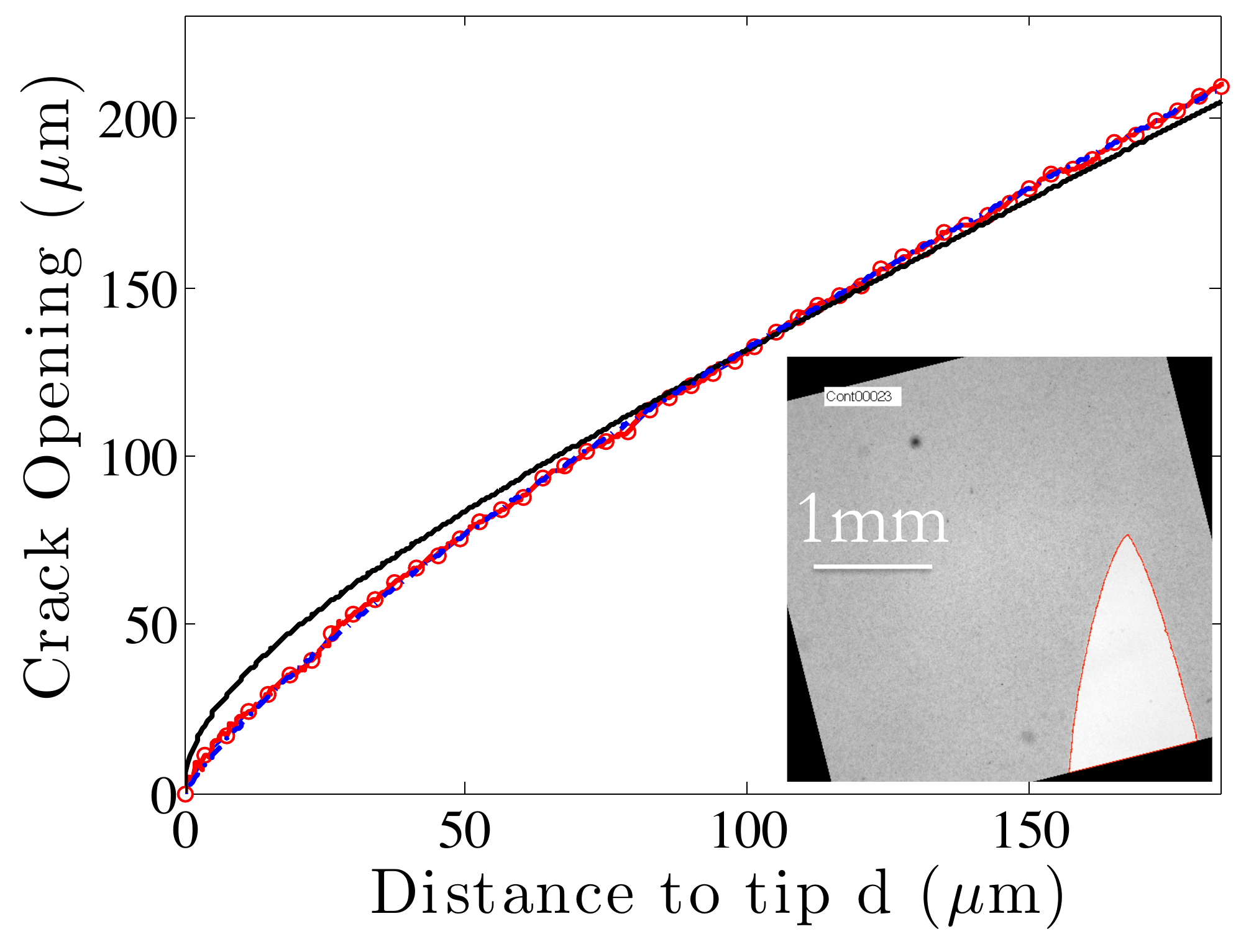}
\vspace{-0.5cm}
\caption{{\bf Rapid crack ($V$=1cm.s$^{-1}$)}. Crack opening (displacement along the crack lips) as a function of the distance d to the crack tip. Red line with circles: experimental results. Black dashed-dotted line: fit of the data with the LEFM prediction \ref{u}, which clearly does not work close to the crack tip; Blue dashed line: fit of the data with the modified Williams expansion \ref{umod}.
{\it Inset}: Picture of a high velocity crack in a dyed 1.5$\% $wt agar gel. The interior of the crack is filled with water. The contour of the crack is superimposed in red. }

\label{crackzoomrap}
\vspace{-0.0cm}
\end{figure}

However, when increasing $V$, the crack shape departs from the LEFM prediction. Fig.~\ref{crackzoomrap} shows the COD for a high velocity crack ($V$=1cm.s$^{-1}$). A video sequence showing the crack propagation can be accessed from the link given in~\cite{SM2}. The crack profile cannot be captured by Williams expansion. Instead of an ``infinite" slope (at the experimental resolution) at the LEFM crack tip, fast cracks propagating in agar gels exhibit a finite slope at the apex with $u(d)$ $\propto$ $d$. A central point of this paper is to determine the very nature of the region close to the crack tip where this is observed. Is it a Dugdale Barenblatt cohesive zone \cite{Lawn}? Is this new crack shape due to the material viscoelasticity $-$ the so-called de Gennes trumpet \cite{Hui1992}$-$? Or is it an effect of large deformations at the crack tip \cite{Long2011}? 
Based on our phenomenological observation, we write a modified Williams expansion series:
\begin{eqnarray}
\tilde{u} (d)=d \left[  {{A}\over{1+({d\over{d^{\star}}}) ^{1/2}}}+\left( {d \over \tilde{d_{1}}}\right) ^{1/2}+ \left( {d \over \tilde{d_{2}}}\right) ^{3/2} \right]
\label{umod}
\end{eqnarray}
where $d^{\star}$ is a crossover length between the linear apex and the classical LEFM expansion. $A$ is a measure of the finite slope at the apex. While the crack opening profile may look like a perfectly linear elastic one at large scale, it is obviously not the case at distances smaller than $d^{\star}$.  Eq.~\ref{umod} fits well the COD over the whole velocity range. $\tilde{d_{1}}$ and $\tilde{d_{2}}$  hardly change with crack velocity. They are both of order  $10^{5} \mu m$ with $\tilde{d_{1}} <  \tilde{d_{2}}$ :  the corresponding terms dominate at distances larger than 1mm.  Each crack propagating at velocity $V$ will be fully characterized by a couple ($A (V)$, $d^{\star}(V)$).

\begin{figure}[H] 
\center
\vspace{0.0cm}
\includegraphics[width=1\columnwidth,height=0.42\columnwidth]{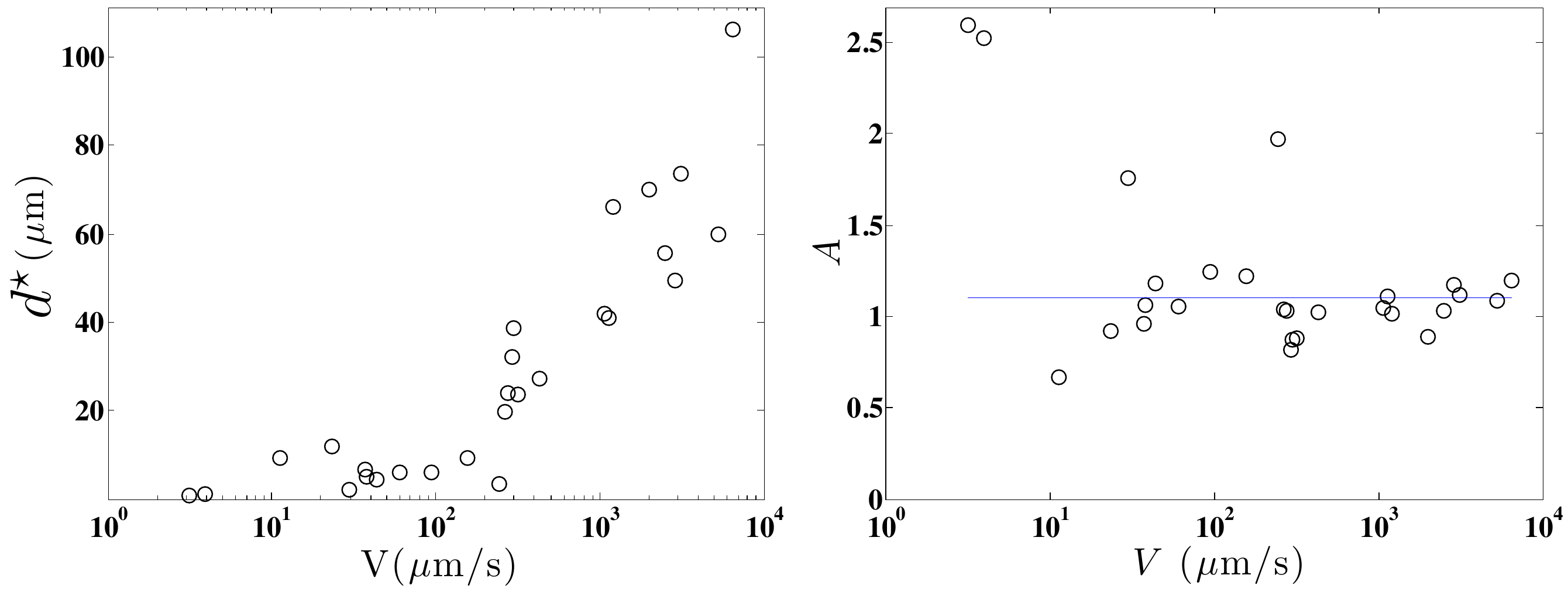}
\vspace{-0.5cm}
\caption{{\bf Parameters of the modified Williams expansion as a function of crack velocity $V$}. {\it Left}. Evolution of $d^{\star}$ with $V$. {\it Right}. Evolution of $A$ with $V$. The blue line is a guide for the eye showing that $A$ is nearly $V$-independent.}

\label{resultsAX}
\vspace{-0.0cm}
\end{figure}

Values of $A$ and $d^{\star}$ are shown in Fig.~\ref{resultsAX}. Except for very low velocities, for which it is difficult to define a linear apex, $A$ appears to be a constant of order unity over the whole velocity range. Meanwhile, a continuous increase of $d^{\star}$ with $V$ can be observed (Fig.~\ref{resultsAX}, left).  $d^{\star}$ appears to be a dynamical lengthscale that vanishes at small velocities and may reach up to 100 $\mu$m at high velocity. Similar results have already been observed by Bouchbinder \& al \cite{Bouchbinder2010} in the context of the dynamic fracture of chemical gels. It is the first time, however, that such an observation is made for quasi-static crack propagation in a soft material.

At distances from the crack tip which are large compared to $d^{\star}$, but for which the $2^{nd} $ term of Eq.~\ref{umod} is still negligible, the dominant contribution to the opening is $\sim$ $A (d^{\star} d)^{1/2}$. By analogy with the Irwin term in Eq.~\ref{u}, we can compute the SIF $K_I (V)$:
\begin{eqnarray}
K_I (V) = A(V) d^{\star} (V)^{1/2} E \sqrt {\pi\over 8}
\label{fittoKI}
\end{eqnarray}
The energy release rate $\cal {G} $ is computed using Eq.~\ref{fittoKI}, and $\cal {G} $$= {K_I}^2/E$ (see Fig.~\ref{results}). The evolution of $\cal {G} $ as a function of $V$ is shown in Fig.~\ref{results}. As expected for such soft materials where viscous processes occur, $\cal {G} $ increases with $V$ due to viscous dissipation within the bulk of the material, or due to dissipative processes at the crack tip.  At low velocity $V$, $\cal {G} $ seems to reach a finite limit $\cal {G} $$_c$ below $0.1 J/{m}^2$. It can be interpreted as the critical energy release rate at the onset of crack propagation. At high velocity, $\cal {G} $ $\gg \cal {G} $$_c $. Increasing $V$ by 4 decades induces more than one order of magnitude increase of $\cal {G} $ . We note that the results obtained for $\cal {G} $ $(V)$ by analyzing crack profiles in agar gels are qualitatively similar to those obtained by integrating force-elongation curves measured during fracture experiments on other biopolymer or physical gels \cite{Seitz2009, Baumberger2009}.

\section{Discussion.}\label{sec4}

A natural assumption would be that $d^{\star}$ is the size of a process zone where viscoelastic dissipative processes take place.  Dugdale$-$Barenblatt models \cite{Lawn} consider a process zone of size $\ell_{DB}$ where cohesive stresses prevent the crack from opening. But, whatever the cohesive stress profile, asymptotic crack opening is predicted to show a cusp in $d^{3/2}$ \cite{Lawn} which does not match our experimental observations at this scale. Besides, as $G'>G"$ over the whole frequency range, the material is elastic at all time scales and there is no reason for the crack tip shape to result from bulk viscoelasticity.

\begin{figure}[H] 
\center
\vspace{0.0cm}
\includegraphics[width=0.95\columnwidth,height=0.70\columnwidth]{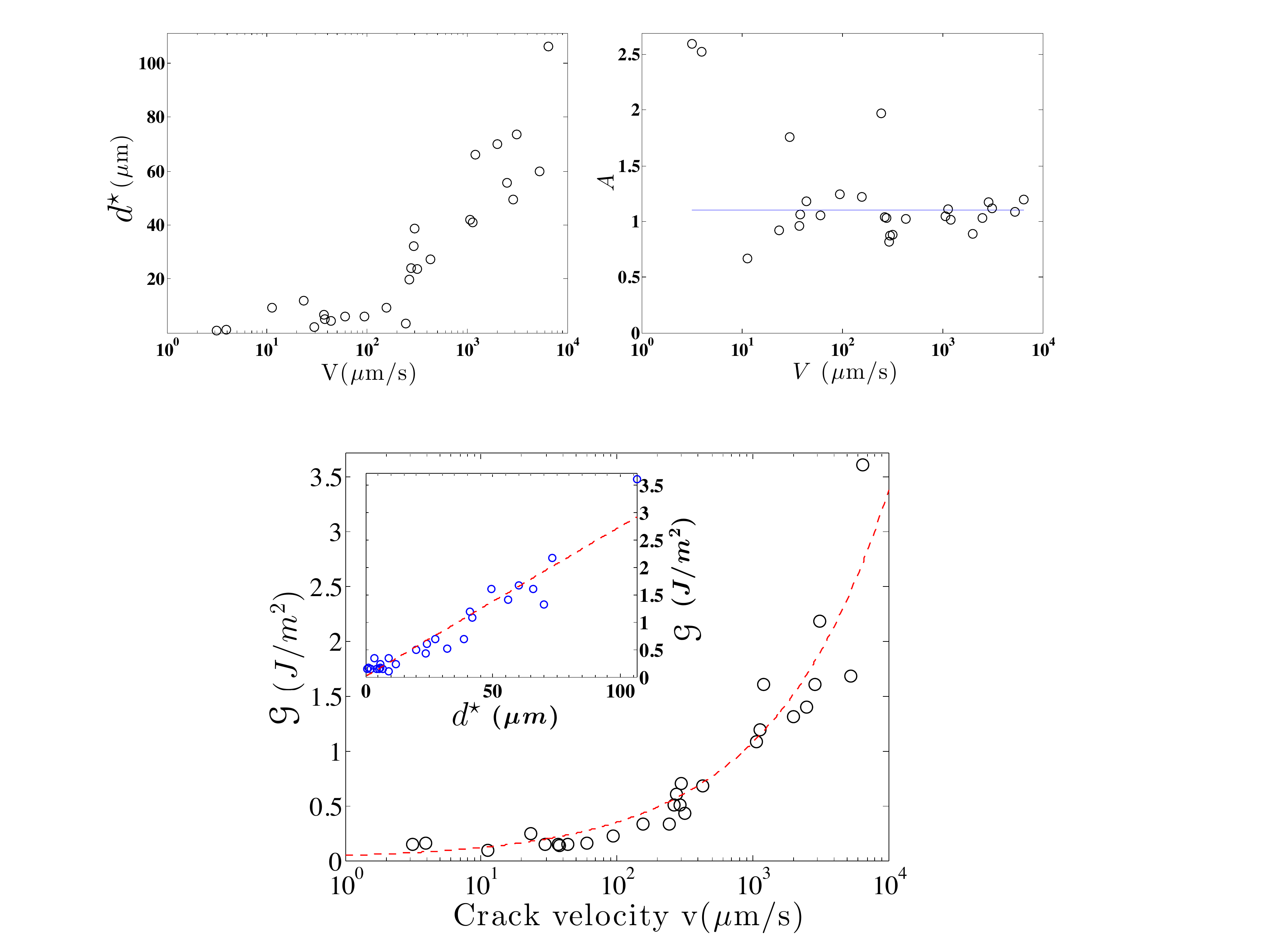}
\vspace{-0.5cm}
\caption{{\bf Evolution of $\cal {G}$ with V}. Black circles are the experimental results computed using Eq. \ref{fittoKI}. The red dashed line is a fit corresponding to $\cal {G}$$=\cal {G}$$_c(1+(V/{\tilde{V}})^{1/2})$. {\it Inset}. Linear variation of $\cal {G}$ with the dynamic lengthscale $d^{\star}$.}

\label{results}
\vspace{-0.0cm}
\end{figure}

Finally, let us assume that  $d^{\star}$ marks a domain of large deformations in the crack tip vicinity, which leads to a strong deviation from LEFM. 

 From Eq.~\ref{fittoKI}, at a distance $r > d^{\star} $ ahead of the crack tip, i.e. within the linear elastic zone, the deformation scales as:
\begin{eqnarray}
\epsilon (r)=\partial _{r} u(r})={K_I (V) \over {E \sqrt {2\pi {r}}}}= { A(V) \over 4 \sqrt {r / d^{\star}}
\label{du}
\end{eqnarray}

$A$ being independent of $V$, with $A \sim 1.2$, we find that $\epsilon (d^{\star})$ is about 30~\%, which is a deformation large enough to justify that LEFM does not apply in this zone. Indeed, Bouchbinder \& al \cite{Bouchbinder2010} have shown that the effect of large deformations becomes measurable within the 10-20\% range. At distances $d<d^{\star}$, the crack tip shape depends strongly on the nonlinear constitutive law of the material. While, for neo-Hookean materials, Geubelle \& Knauss \cite{Geubelle1993} predict a ''linear-like" parabolic crack tip which was observed in \cite{Livne2013}, Long \& al compute the non linear crack tip shape in an exponentially stiffening material \cite{Long2011}: they predict a wedge-like shape similar to what is observed in our case.

We now discuss the $V$-dependence of $\cal {G}$ which can be written as:
\begin{eqnarray}
{\cal G} ={\cal G} _c+ {\cal G} _{vis} (V)
\label{Gv}
\end{eqnarray}
where ${\cal G} _c$ is the critical energy release rate (fracture energy at vanishing crack velocity) and ${\cal G} _{vis} (V) $ is the rate-dependent energy (per unit crack surface created) dissipated during steady-state propagation at velocity $V$. 

Following \cite{Bouchbinder2011a}, we estimate the bulk dissipation for a stationnary crack (for which time derivatives can be replaced by space derivatives, i.e. $V \partial t =  \partial r$) propagating at vanishing velocity $V$ in the following way:
\begin{eqnarray}
{\cal G}_{vis} & = &{\eta \over{V} } \int (\partial _{r} \dot{u})^2 r dr \\
& = & \eta V \int_{r =d_c}^{r=\infty} (\partial _{rr} u)^2 r dr
\label{Gvis}
\end{eqnarray}
where we have not taken into account the angular variation of $u$ since, once integrated in Eq.\ref{Gvis}, it will only modify the prefactor.
In  \cite{Bouchbinder2011a},  $\eta$ is an effective material viscosity. $d_c$ is the microscopic length scale where the linear theory fails, which is actually the lengthscale $d^{\star}$ measured in our experiment.

In this perturbative approach, one can use the elastic asymptotic crack displacement field $u(r) \propto (K _{I C}/E) \sqrt {r} $, and $\partial _{rr} u\propto -K _{I C}/E r^{-3/2}$; Hence:
\begin{eqnarray}
{\cal G}_{vis} \propto  {{\cal G} _c \eta  \over E d^{\star }(V)}V
\label{Gvis2}
\end{eqnarray}

\noindent where ${\cal G} _c=K _{I C}^2/E$.

Computing ${\cal G}_{vis} (V)$ at higher velocity is a more delicate task. Indeed, at finite $V$, local dissipation at a distance $r$ from the crack tip depends on the local deformation rate at pulsation $\omega = V/r$ \cite{Persson2005}. Moreover, the perturbative approach, valid when ${\cal G}_{vis} \ll {\cal G}_c$ does not hold anymore when ${\cal G} \gg {\cal G}_c$.  If we recall that $\epsilon^{\star}= \epsilon (r=d^{\star}(V))$ is velocity-independent, then Eq.~\ref{du} directly leads to:
\begin{eqnarray}
{\cal G}(V)={{K_I }^2(V) \over E} = 2\pi E(\epsilon^{\star})^2  d^{\star}(V) 
\label{Gstar}
\end{eqnarray}

{\it i.e.}  ${\cal G} (V) /E \propto d^{\star}$. The inset of Fig.~\ref{results} shows ${\cal G}$ versus $d^{\star}$. A linear relation ${\cal G} (V)= \alpha d^{\star}$ fits indeed the data with $\alpha= 2.7~10^4 Pa$ .
For large $V$, as ${\cal G}  \sim {\cal G} _{vis}$, we have: 
\begin{eqnarray}
{{\cal G} _{vis}(V) \over {\cal G}_c}={d^{\star}(V) \over d^{\star}(V=0)}
\label{ratio}
\end{eqnarray}

Matched asymptotics enables to eliminate $d^{\star}(V)$ in Eq.~\ref{Gvis2} using Eq.~\ref{ratio}, and we get, for high velocities:
\begin{eqnarray}
{\cal G}_{vis} \sim  {\cal G} _c \sqrt {\eta V \over E d^{\star}(V=0)}
\label{Gvis3}
\end{eqnarray}
Over the whole velocity range, we thus have: 
\begin{eqnarray}
{\cal G} \sim {\cal G}_c(1 +  \sqrt {\eta V \over E d^{\star}(V=0)})={\cal G}_c (1+ \sqrt {V \over \tilde{V}})
\label{Gvis4}
\end{eqnarray}
where \begin{eqnarray}
\tilde{V}= {Ed^{\star}(V=0) \over\eta}
\label{vtild}
\end{eqnarray}
Agreement between the measured fracture energy ${\cal G}(V)$ and Eq.~\ref{Gvis4}  is good within the explored velocity range (Fig.~\ref{results}). 

The resulting value for the critical energy release rate ${\cal G}_c = 0.01 J/m^2$ is reasonable. 
Let us recall that ${\cal G}_c$ is the surface energy needed to break the bonds crossing the fracture plane. Given the weakness of H bonds compared to the covalent bonds, fracture of these physical gels is reported to occur via unzipping of the junctions of energy $ U_H $ \cite{Baumberger2006a}. The model of Lake \& Thomas (1967), stating that all the energy stored in a chain containing $n$ residues is dissipated at rupture, ${\cal G}_c$ should scale like $U_H / {\xi a}$ where $\xi$ is the distance between the junctions and $a$ is the size of a residue. With $U_H=0.1 eV$, $\xi = 10nm$ and $a = 0.3nm$, we find ${\cal G}_c = 5.10^{-3} J/{m}^2$ which is in good agreement with the value measured thanks to our model.

The fit also gives a value for $\tilde{V} = 0.1 \mu m.s^{-1}$ which is below the minimum crack velocity that we could reach. For a crack of velocity $\tilde{V}$, ${\cal G}$ is expected to be twice  ${\cal G}_c$. What we observe experimentally is in fact an intermittent crack propagation regime. At these velocities, we suspect that a moderate stress relaxation of the network is sufficient to dissipate elastic energy and to decrease ${\cal G}$ below its critical value ${\cal G}_c$ causing the crack to grow intermittently. 
From Eq.~\ref{vtild}, we get $\eta=E d^{\star}(V=0)/\tilde{V}$. When $V{\rightarrow}0 $, no large deformation is expected at the crack tip and the limitation to LEFM is thus the extent of the Dugdale$-$Barenblatt cohesive zone~$\ell_{DB}= {\cal G}_c/\sigma_y={\cal G}_c/E \simeq 0.1\mu m$. Consequently, $\eta$ is found to be $E\ell_{DB}/\tilde{V}\sim 6.10^4 Pa.s$. 

 We can wonder how a material with rate independent $G'$ and $G"$ exhibit rate-dependent fracture properties. The answer lies in the rate-dependency of the highly non linear processes occuring in the vicinity of the crack tip. One of them is viscous chain pull-out, proposed by Baumberger in the case of gelatin \cite{Baumberger2009}. It results in an effective viscosity scaling like $ \eta_S \Lambda /{\xi}=20 \eta_S$ where $\eta_S=10^{-3} Pa.s$,  is the solvent viscosity because we consider here chain/water friction. For our agar gel, this effective viscosity is orders of magnitude too low. 

Another option is to consider that dissipation stems from the dynamics of agar chain segments confined in the junction zones. Indeed, reptation processes result in an effective viscosity $\eta_{eff}=G'\tau_{rep}$\cite{rubinstein} where $\tau_{rep}$ is the chain reptation time. In our case, the reptation time is the time at which our biopolymer network relaxes. At our working concentration, \cite{Labropoulos2001} found that agar networks relax on timescales of $10^2-10^3s$. This would result in $\eta_{eff}=10^7-10^8 Pa.s$ which is now far larger than our measurement and confirms that our material does not flow on experimental timescales. But if we now consider the material in the vicinity of the crack tip, stress-aided reptation of the strongly deformed chains may significantly decrease $\tau_{rep}$ by 1 or 2 orders of magnitude \cite{Lefrancthese}, which leads to $\eta_{eff}$ in the measured range.

Finally, the observed increase of $d^{\star}$ with $V$ suggests an increase of the deformation to failure for increasing crack velocity and confirms the rate-dependency of the failure properties of agar networks (see for example Fig. 5, 6 and 7 in~\cite{Mcevoy1985}). Below a velocity of order $\tilde{V}$ (which is out of our experimental range), stress-aided partial unzipping of the junctions leads to an increased strain to failure $\epsilon_R$ of the network. At moderate $V$, $\epsilon_R$ reaches a mininum lower than $10\%$: Fracture occurs in the linear elastic regime, and $d^{\star}$ is thus very small. Finally, at higher rates, $\epsilon_R$ increases with $V$, and reaches values larger than $10\%$. This is the reason why we observe a more extended zone of high strains in the crack tip vicinity when $V$ is much larger than $\tilde{V}$. A quantitative understanding of the rate depency of polymer gel failure properties at high deformation rates still lacking.

\section{ Conclusion.}

In our experiments, we were able to recover the energy release rate measured by other authors \cite{Baumberger2006a, Properties2000, Tanaka2000} from macroscopic experiments, by analyzing the morphology of the crack tip.

Because fracturing our physical gels at high velocity $V$ causes a great deal of dissipation, the stress intensity factor $K_I$ must increase with $V$ accordingly. This induces large deformations in the crack tip vicinity, which translates into an earlier departure from LEFM ($d^{\star}$ increases with $V$). A full understanding of the very shape of the crack at small scales is still in progess, based on the work of Livne \& al \cite{Livne2013} and Long \& al \cite{Long2011}.

We will take advantage of our new experimental setup to study less concentrated gels with lower moduli, in order to observe the emergence of the cohesive zone. More fundamentally, the growth and possible divergence of the size of the process zone in the vicinity of the sol gel transition will be studied.
In the future, these experiments can be extended to other types of liquid to solid transitions such as the colloidal glass transition.

\section{Acknowledgements.} We are indebted to Fabrice Monti and Patrick Tabeling for introducing us to microfluidics and for their help all along the project. Enlightening discussions with Eran Bouchbinder, Olivier Dauchot, Guylaine Ducouret, Fran\c cois Hild, Laurent Ponson and Matteo Ciccotti are also acknowledged. We also thank Jean-Philippe Bouchaud for his constant support and his careful reading of the manuscript. Our research was funded by Agence Nationale pour la Recherche (ANR) through the F2F project.

\section*{References}

\bibliography{maximeqat.bib}

\end{document}